\documentclass[aps,prl,floatfix,12pt,showpacs]{revtex4}

\usepackage{epsfig}

\begin{document}

\vspace{0.4cm}

\title{
    Sequence of first-order quantum phase transitions in a           
    frustrated spin half dimer-plaquette chain                
}
\author{J. Schulenburg and J. Richter }
%\address
\affiliation{
Institut f\"ur Theoretische Physik, Universit\"at Magdeburg, P.O.Box 4120, 
D-39016 Magdeburg, Germany
}

\begin{abstract}
We study
the frustrated dimer-plaquette quantum spin chain for ferromagnetic
dimer bonds. This quantum system undergoes a series of 
first-order ground-state phase transitions driven by
frustration or by a magnetic field.
We find that the different nature of the ground-state phases has a strong 
influence on the magnetization curve as well as the low-temperature
thermodynamics. In particular, the magnetization curve 
exhibits plateaus and jumps, the number of which depends on the strength of
frustration.
The temperature dependence of the susceptibility may either show 
activated spin-gap behaviour for small frustration
or Curie like paramagnetic behaviour for strong frustration.   
\end{abstract}
\pacs{75.10.Jm}

\date{June 27, 2002}
\maketitle

\section{Introduction}
Over the last years much attention  
has been concentrated on  
the physics of low-dimensional quantum spin systems. 
In particular, zero-temperature
phase transitions and exotic magnetization curves 
in frustrated quantum magnets are in the focus of
investigations.     
Besides continuous quantum phase-transitions also 
remarkable first-order transitions can be driven 
by frustration.
The theoretical studies  have benefitted from recent
experimental results on low-dimensional spin-half 
antiferromagnets like 
quasi-onedimensional ladder, zigzag and spin-Peierls 
systems 
as well as gapped quasi-twodimensional
quantum magnets like $CaV_4O_9$  
\cite{taniguchi95,ohama97} and $SrCu_2(BO_2)_3$ 
\cite{kageyama99,ueda99,onizuka00}.

\begin{figure}
\begin{center}
 {\unitlength=1.25mm \input{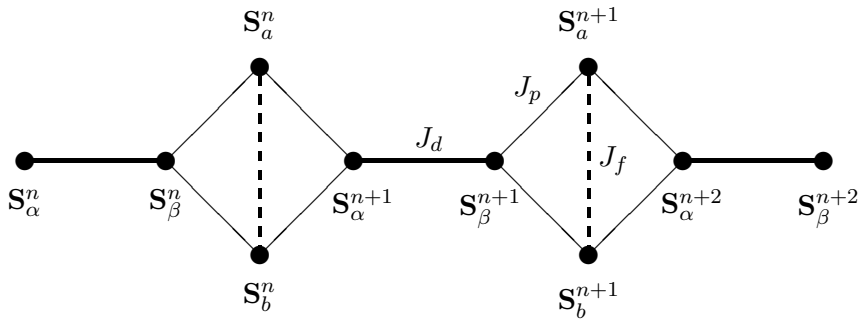}}
\end{center}
\caption{
  The fustrated dimer-plaquette chain (FDPC).
}
\label{fig1}
\end{figure}

Among the various models for low-dimensional quantum 
magnets the  spin-half 
frustrated dimer-plaquette chain (FDPC) has been  
discussed in a series of papers in
recent years. This FDPC was introduced in \cite{ivanov}   
and its Hamiltonian reads (see also Fig. \ref{fig1})
\begin{equation} \label{eq1}
H = H_{dp}+ H_f=
J_d \sum^{N_p}_{n=1}{{\bf S}^n_{\alpha}{\bf S}^n_{\beta}}
+  J_p \sum^{N_p}_{n=1}
%{({\bf S}^n_{\beta}{\bf S}^n_{a}+{\bf S}^n_{\beta}{\bf S}^n_{b}+
%{\bf S}^n_{a}{\bf S}^{n+1}_{\alpha}+{\bf S}^n_{b}{\bf S}^{n+1}_{\alpha})}.
 ({\bf S}^n_{a}     + {\bf S}^n_{b})
 ({\bf S}^n_{\beta} + {\bf S}^{n+1}_{\alpha})
+  J_f \sum^{N_p}_{n=1}
{{\bf S}^n_{a}{\bf S}^n_{b}}
\end{equation}
where $N_p=N/4$ is the number of plaquettes (unit cells)
and $N$ is the number of spins. We consider finite chains of $N$ spins  with
periodic boundary conditions.
The Hamiltonian fulfills the important relation 
$[H,({\bf S}^n_{a}     + {\bf S}^n_{b})^2]_-=0$. Hence 
the vertical dimer  spins  
${\bf S}^n_{a}$ and ${\bf S}^n_{b}$ form a composite 
spin ${\bf S}_{ab}^n=({\bf S}^n_{a}     + {\bf S}^n_{b})$ with eigenvalues 
$({\bf S}^n_{ab})^2=
S^n_{ab}(S^n_{ab}+1)$ and $S^n_{ab}=\{1,0\}$.    
The physics of this model was discussed in the literature for  
antiferromagnetic coup\-lings $J_d, J_p, J_f \ge 0$
\cite{ivanov,richter,koga00,schul02,koga02}. 

In the  unfrustrated version ($J_f=0$, $H = H_{dp}$) 
the model may serve 
as the
onedimensional counterpart of the 1/5-depleted 
square lattice of $CaV_4O_9$ \cite{katoh95,ivanov,richter}.
In this case ($J_f=0$) all composite spins have eigenvalues
$S_{ab}^n=1$ in the singlet ground state as well as in the first 
triplet excitation and the ground-state physics of (\ref{eq1})
is determined by the competition of dimer ($J_d$) and 
plaquette ($J_p$) bonds. 
In the limit  $J_d \gg J_p$ the low-energy physics of the model 
is that of the  Haldane
chain with an effective coupling $J_{eff}=J_p^2/2J_d$, whereas  for 
$J_p \gg J_d$ a plaquette phase is realized \cite{richter}.  
The singlet-triplet excitation gap is always
finite for $J_p > 0$.
A special feature of the model is the existence of a class of exact
product eigenstates, for which  the composite spins  of certain  
plaquettes form 
a vertical dimer singlet (i.e. $S^n_{ab}=0$ for certain $n$).
The finite strips (fragments) 
between two vertical dimer singlets contain
vertical dimer triplets $S^{n'}_{ab}=1$ and are  decoupled from each 
other. Therefore these product eigenstates correspond to a fragmentation
of the chain. 

 Taking into account frustration $J_f > 0$ the formation of vertical
dimer singlets becomes  energetically more favourable and the 
system undergoes at a 
critical frustration
$J_f^{crit} =f(J_d,J_p)$ a first-order quantum phase 
transition from the collective ground state with $S^n_{ab}=1$ for all $n$ 
to a  fragmented  dimer product ground state 
\begin{equation} \label{eq2}
 {}|\Psi_{0\ldots 0}\rangle =
 \prod_{n=1}^{N_p} 2^{-\frac{1}{2}}\left(| \uparrow_a^n\downarrow_b^n \rangle
                       -|\downarrow_a^n\uparrow_b^n \rangle\right)
 \prod_{n=1}^{N_p} |d^n_{\alpha \beta}\rangle ,
\end{equation}
with
$S^n_{ab}=0$ for all $n$. 
For $J_d > 0$ the dimer state $|d^n_{\alpha \beta}\rangle$ )
is a singlet and 
the energy of 
$|\Psi_{0\ldots 0}\rangle$  is $E/N_p=-3J_f/4-3J_d/4$.

In the special limit  $J_f=J_d$ the critical value is $J_f^{crit}=1.2210 J_p$
\cite{koga00}. In this limit the FDPC has a close relation to the
two-dimensional spin model for SrCu$_2$(BO$_3$)$_2$ \cite{ueda,koga00,schul02} 
and is called orthogonal-dimer chain. 
Similar to the quasi-twodimensional spin system  SrCu$_2$(BO$_3$)$_2$
\cite{kageyama99,onizuka00}
the FDPC exhibits nontrivial magnetization plateaus \cite{koga00,schul02}.
A novel property of the FDPC, not found so far in quantum spin systems, is 
the existence of an infinite sequence of plateaus \cite{schul02}.

Very recently, the FDPC has been considered for arbitrary spin quantum numbers $s \ge
1/2$ and a series of $2s$ 
first-order quantum phase transitions has been found by Koga and Kawakami
\cite{koga02}. Koga and coworkers argue that the first-order
phase transitions described by the FDPC possess most of the essential
features inherent in frustrated quantum spin systems \cite{koga00,koga02}. 
In that sense the FDPC may serve as
a prototype model for spin sytems showing first-order  
quantum phase transitions. 

In the present paper we extend the model to
ferromagnetic bonds $J_d$. We notice, that the fragmentation 
of the chain due to 
vertical dimer singlets occurs for $J_d < 0$, too.
In particular, 
the  dimer product state  (\ref{eq2}) is an eigenstate of (\ref{eq1}) also
for $J_d<0$, however, the dimer state $|d^n_{\alpha \beta}\rangle$ in 
$|\Psi_{0\ldots 0}\rangle$ is a triplet. 
The energy of $|\Psi_{0\ldots 0}\rangle$  is then $E/N_p=-3J_f/4 + J_d/4$.

\section{Ground state phase diagram} 
We start from the energy eigenvalues $E$ of model (\ref{eq1}) 
and write the dependence of $E$ on $J_f$ in explicit form 
\begin{equation}
E(J_p,J_d,J_f) =E_{dp}(J_p,J_d) 
+ J_f \left (\frac{1}{4} N_t - \frac{3}{4} N_s\right ),
\end{equation} 
where $N_s$ is the number of vertical dimers with 
composite spin $S^n_{ab}=0$ and
$N_t$ is the number of vertical dimers 
with composite spin $S^n_{ab}=1$. We have $N_s+N_t=N_p$.
Since every vertical dimer singlet leads to an energy  gain
of $J_f$ in the second part of the energy the frustration favours states
with $S_{ab}^n=0$, i.e. a fragmentation of the chain.   
The lowest state of these finite fragments  
has a total spin  $S_{frag}=1$ for $J_d<0$.
We define the length $k$ of a fragment, 
as the number  of triplet composite spins
$S^n_{ab}=1$ between two singlet composite spins.

\begin{figure}
 \begin{center}
  \epsfig{file=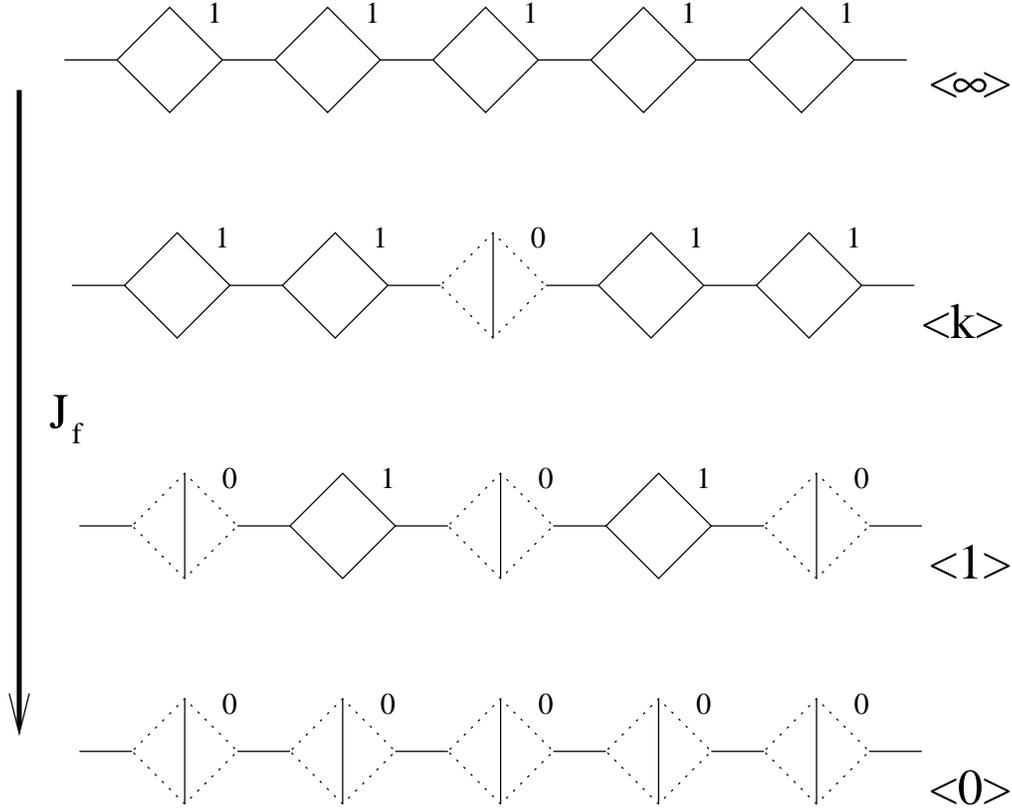,scale=1.0}
 \end{center}
\caption{ \label{dp_frag}
Change of the ground state with increasing frustration $J_f$. For small
fustration $J_f$ all composite spins are in the triplet
state and 
there is no fragmentation.
Increasing $J_f$ some composite spins may prefer a vertical  
singlet state (indicated by
vertical lines) and the chain may split into identical fragments of length
$k$ (the spin-spin correlation is zero along the $J_p$ bonds neighbouring a
vertical singlet as indicated by
dotted lines).
For large $J_f$ all composite spins are in the singlet state and the  
dimer product state (\ref{eq2}) is the  ground state. 
}
\end{figure}

According to the so-called linear programming scheme the extrema of the
energy belong to states with fragments of identical length $k$
\cite{niggemann}.  In what follows we use the notation $<\!\!k\!\!>$ for a
fragmented state consisting of identical fragments of length $k$. 
The dimer product state (\ref{eq2}) corresponds to fragments of length
$k=0$, i.e. $|\Psi_{0\ldots 0}\rangle= <\!\!0\!\!>$.
It is evident that the system undergoes at least one transition 
from the nonfragmented ground state  $<\!\!\infty\!\!>$ with 
$S^n_{ab}=1$ ($n=1 \ldots N_p$) at small $J_f$ to the dimer product 
ground state $<\!\!0\!\!>$
with 
$S^n_{ab}=0$ ($n=1 \ldots N_p$) for large $J_f$ (see Fig. \ref{dp_frag}). 
As mentioned above, for antiferromagnetic $J_d$ indeed we have a direct
transition $<\!\!\infty\!\!> \to <\!\!0\!\!>$.  
However, for ferromagnetic $J_d$ the question for
the existence of 
intermediate  ground states $<\!\!k\!\!>$ 
with fragments of finite length $0 < k < \infty$
needs more attention.

We start from the ground state $<\!\!\infty\!\!>$ for low frustration. The
energy per unit cell  of this state is 
\begin{equation} \label{eq3}
 e^f_{\infty}=E^f_\infty/N_p=e_{\infty}+J_f/4 ,
\end{equation}
where $e_{\infty}$ is the energy of the ground state for the unfrustrated
chain ($J_f=0$).
We determine $e_{\infty}$ by finite-size extrapolation 
of chains of length $N=8,16,24,32$ as well as
perturbation theory. 
The energy per unit cell $e^f_k$ of a fragmented state $<\!\! k \!\!>$ is given by 
the sum of the energies $E_k+kJ_f/4$ of the fragments  and of the energies 
$3J_f/4$ of the singlets separating the fragments
\begin{equation} \label{eq4}
 e^f_k
         =\frac{E_k+(k-3)J_f/4}{k+1}.
\end{equation}
$E_k$ is the  energy of the unfrustrated ($J_f=0$) fragment of length
$k$ consisting of  $k$ plaquettes and $k+1$ dimer bonds $J_d$ and corresponds 
to the  
energy of a finite  dimer-plaquette chain with $4k+2$ spins
and open boundary conditions.

Using Lanczos algorithm we can exactly caculate the energy of the chain
fragments $E_k$ up to $k=8$ 
(i.e. $N=34$).
For larger fragment lengths
and for the  state  $<\!\!\infty\!\!>$ we need approximations. 
In the limits  $|J_d|/|J_p|\ll 1$ and 
$|J_p|/|J_d|\ll 1$ perturbation theory is appropriate since for $J_d=0$ and
$J_p=0$ the ground states are known as simple product states. 
We use here Rayleigh-Schr\"odinger perturbation theory up to 3rd order. 
The comparison of perturbation theory with Lanczos results shows that the
perturbation theory yields reliable results even for 
$|J_d/J_p|\approx 1$.
\begin{figure}
 \begin{center}
 \epsfig{file=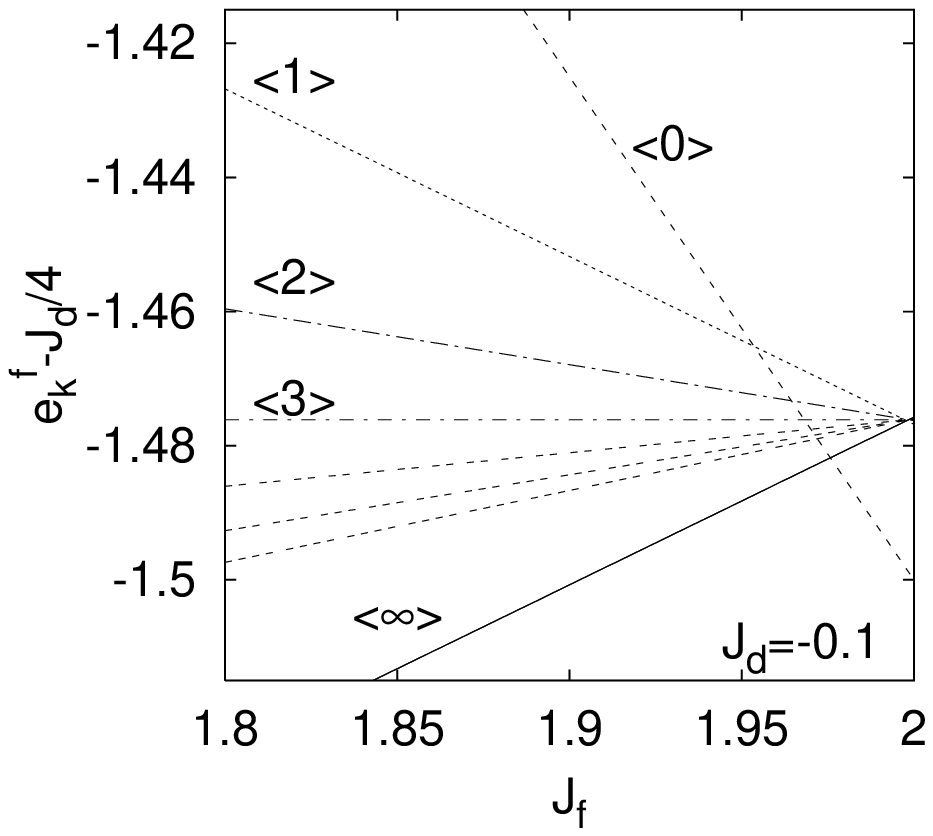,scale=0.75}
 \epsfig{file=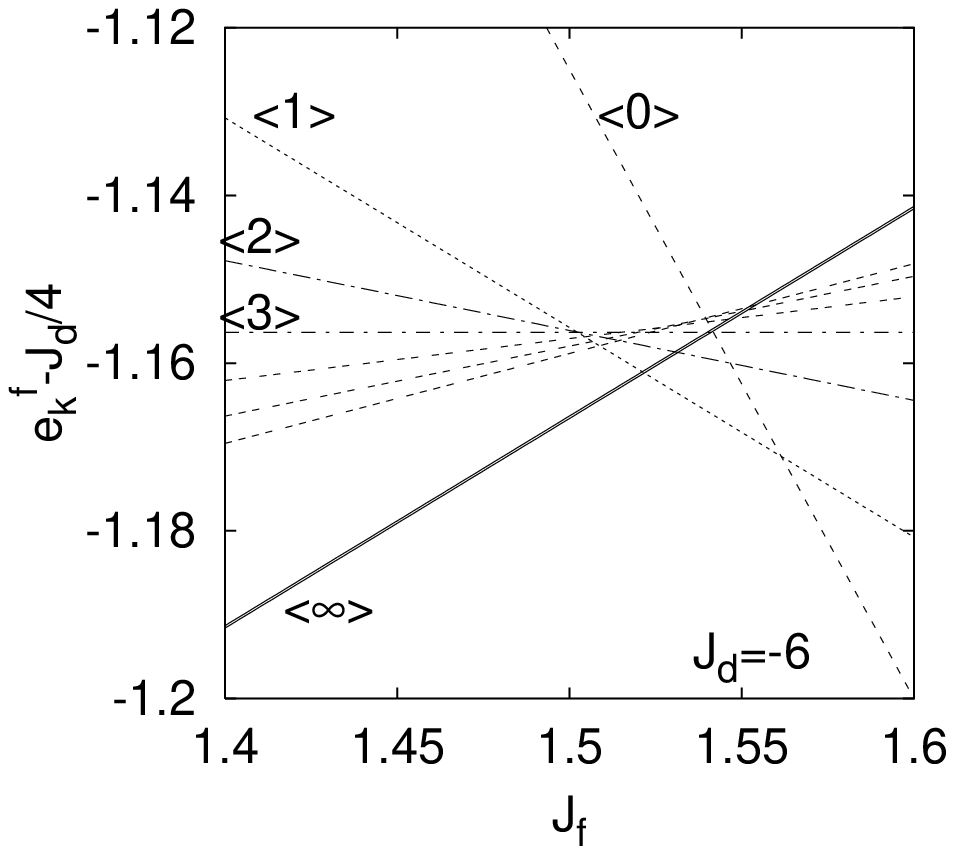,scale=0.75}
 \epsfig{file=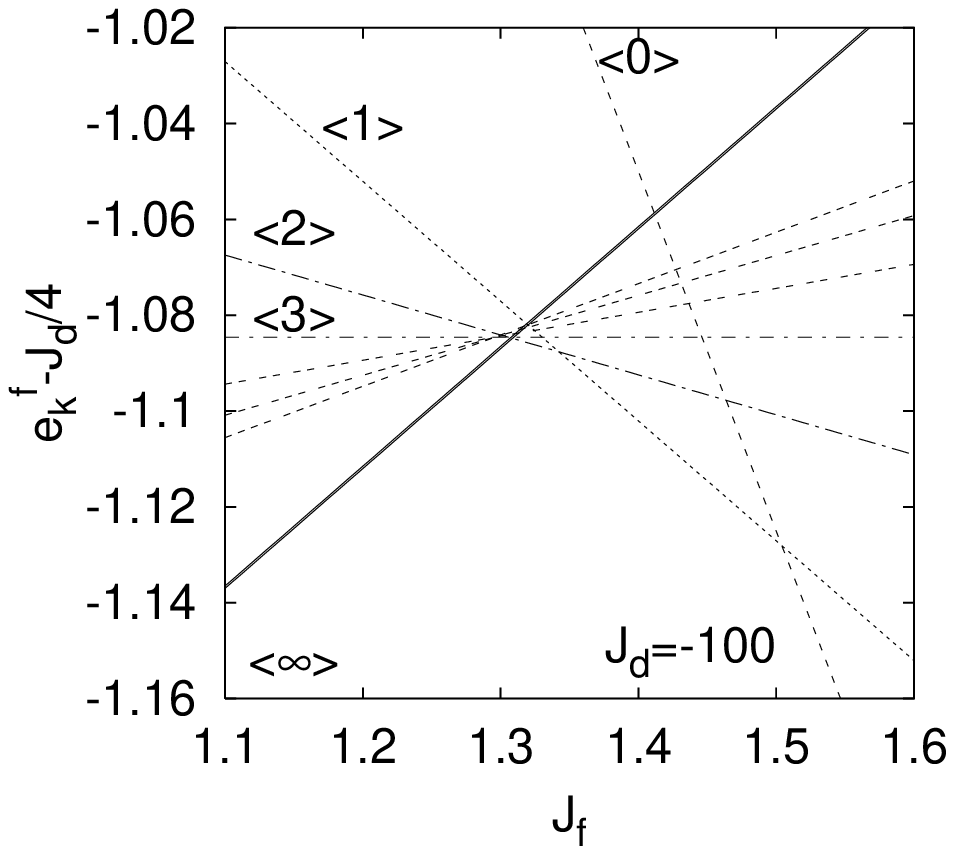,scale=0.75}
 \epsfig{file=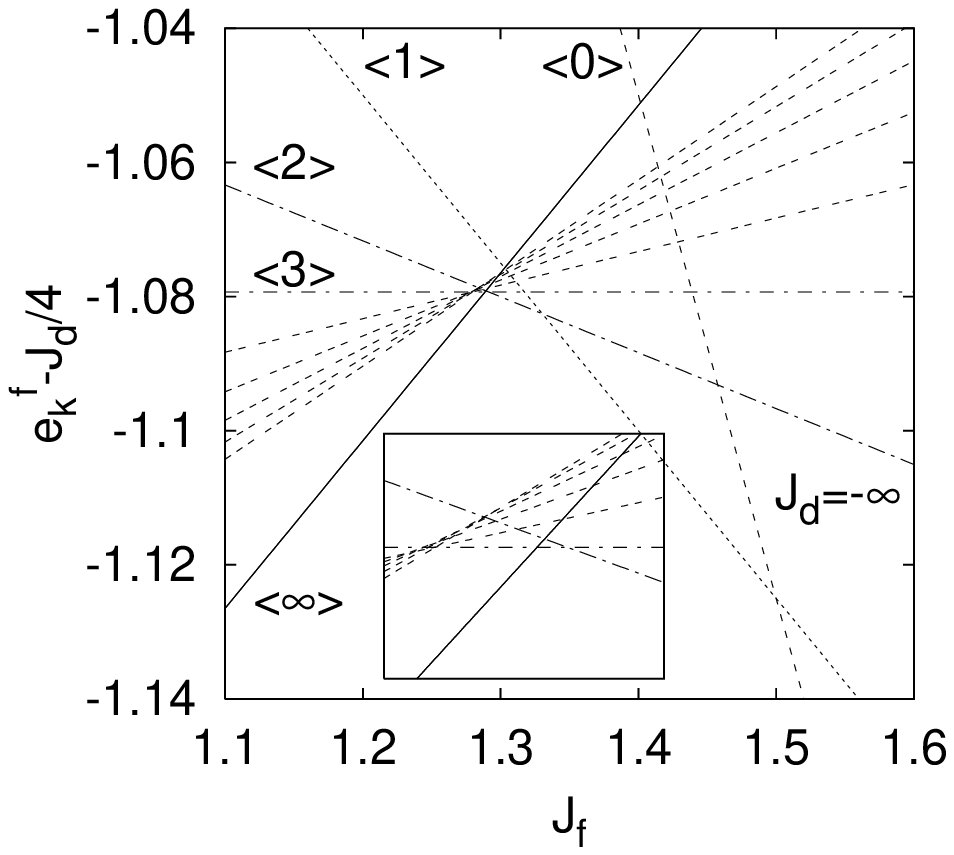,scale=0.75}
 \end{center}
\caption{ \label{dpejf}
  Energies $e_k^f$ (Eq. (\protect{\ref{eq4}}))
 versus 
 $J_f$ for various strengths of ferromagnetic
 dimer bonds ($J_d=-0.1$, $J_d=-6$, $J_d=-100$ and $J_d=-\infty$) and 
 fixed $J_p=1$. 
For better comparison we have subtracted 
$e(J_p=0,J_f=0)=J_d/4$ from $e_k^f$.
 Curves are presented for $k=0,\ldots,7$ (exact results) and for
$k=\infty$ (finite-size extrapolation). The accuracy of the 
$k=\infty$ results corresponds to the thickness of the solid line.
We have labeled by 
$<\!\!k\!\!>$ only the most relevant curves with $k=0,1,2,3,\infty$.
The inset in the figure for $J_d =-\infty$ shows the transition region
between $<\!\! \infty \!\!>$, $<\!\! 3 \!\!>$ and $<\!\! 2  \!\!>$ with an
enlarged scale.
}
\end{figure}
Fig. \ref{dpejf} shows the lowest energies of the state  $<\!\!k\!\!>$
with $k=0,1,2,3,\ldots$ und $k=\infty$ versus $J_f$ for several $J_d$.
The intersection points of the lowest lines 
determine the positions of first-order quantum phase transitions.
It becomes obvious that the number of phase transitions changes from one 
to four increasing the strength of ferromagnetic $J_d$.  
The existence of further transitions can be excluded by analyzing the
$k$-dependence of  $E_k$ according to Niggemann et al. \cite{niggemann}.
The transition point $J_f^{k_1,k_2}$
between two ground-state phases $<\!\! k_1 \!\!>$ and
 $<\!\! k_2 \!\!>$
are obtained by the relation 
$e^f_{k1}=e^f_{k2}$. 
Using Eqs. (\ref{eq3}) and (\ref{eq4}) we find
\begin{equation}
J_f^{k,\infty}=E_k-(k+1)e_{\infty}  \qquad ; \qquad
     J_f^{k-1,k}=(k+1)E_{k-1}-(k)E_{k}.
  \label{dpjfcrit}
\end{equation}
The corresponding phase diagram is shown in Fig. \ref{dpfmphase}. 
\begin{figure}
 \begin{center}
  \epsfig{file=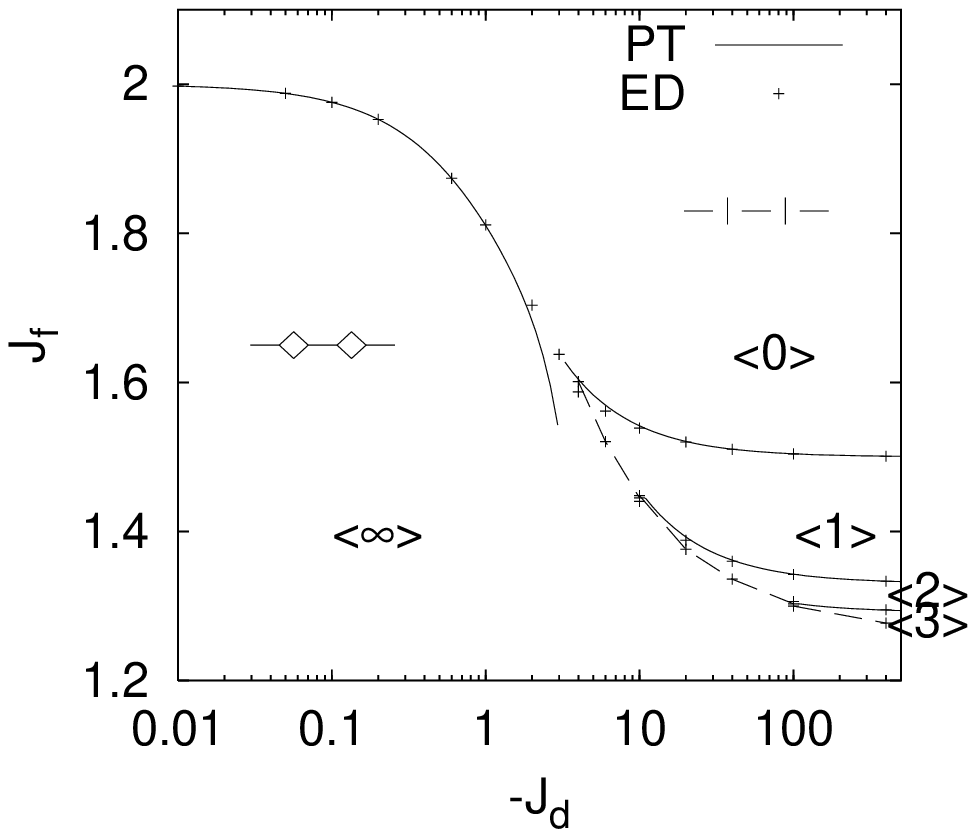,scale=1.0}
 \end{center}
\caption[Grundzustands-Phasendiagramm $J_d>0$]{ \label{dpfmphase}
 Ground-state phase diagram for 
ferromagnetic $J_d<0$ and fixed $J_p=1$ (see text).\\
solid line - perturbation theory, crosses - exact diagonalization
}
\end{figure}
\begin{figure}
 \begin{center}
  \epsfig{file=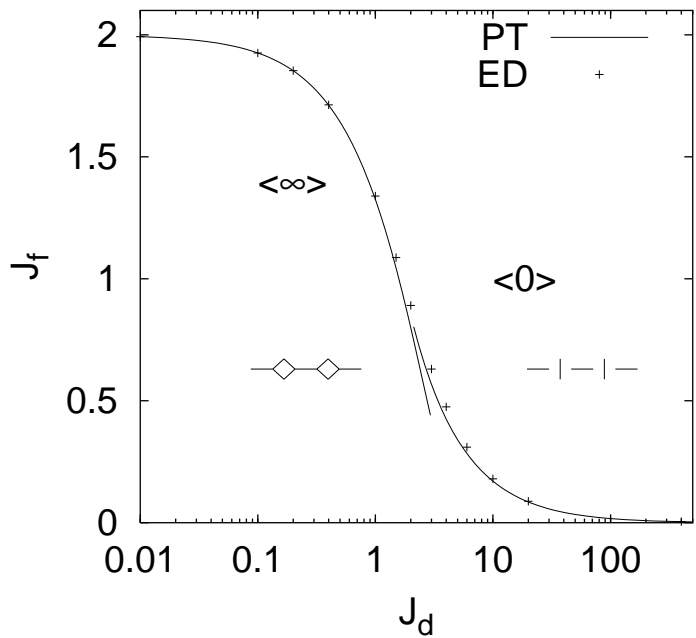,scale=1.0}
 \end{center}
\caption[Grundzustands-Phasendiagramm $J_d>0$]{ \label{dpafphase}
 Ground-state phase diagram for 
antiferromagnetic $J_d>0$ and fixed $J_p=1$ (see text).\\
solid line - perturbation theory, crosses - exact diagonalization
}
\end{figure}
For completeness and comparison we have reconsidered the case of  
antiferromagnetic 
$J_d$ \cite{ivanov,richter}
including perturbation theory and an enlarged exact-diagonalization 
data set, see Fig. \ref{dpafphase}.
For small strength  of dimer coupling $|J_d|\lesssim 3$ 
the situation is similar for both ferro- and antiferromagnetic 
$J_d$, i.e. we have a direct transition 
$<\!\!\infty\!\!> \to <\!\!0\!\!>$. Increasing the strength of 
ferromagnetic dimer coupling 
to $J_d \approx -4$ the situation is changed and we find an additional
intermediate phase $<\!\!1\!\!>$. Further increasing 
$|J_d|$ we find another intermediate phase  $<\!\!2\!\!>$, and finally we
have a sequence of transitions $<\!\!\infty\!\!> \to <\!\!3\!\!> \to
<\!\!2\!\!> \to <\!\!1\!\!> \to <\!\!0\!\!>$ at very strong ferromagnetic
$J_d$. In this extreme limit $J_d \to -\infty$  
the FDPC can be mapped onto the
frustrated diamond chain and 
our results are in agreement with the results reported in Ref. 
\cite{niggemann}.

\section{Influence of a magnetic field on the ground-state phases}
As discussed in Refs. \cite{koga00,schul02}
the FDPC with antiferromagnetic $J_d$ exhibits nontrivial magnetization 
plateaus.
Therefore we consider now 
the Hamiltonian (\ref{eq1}) including an external magnetic field 
\begin{equation} \label{eq5}
H = H_{dp}+ H_f - \; h S^z_{tot},
\end{equation}
where the z-component of the total spin is given by $S^z_{tot} = \sum_i
S^z_i$. 
For ferromagnetic $J_d$ we have a more complex zero-field 
ground-state phase diagram and 
we can expect interesting effects caused by a magnetic field.
\begin{figure}
 \begin{center}
  \epsfig{file=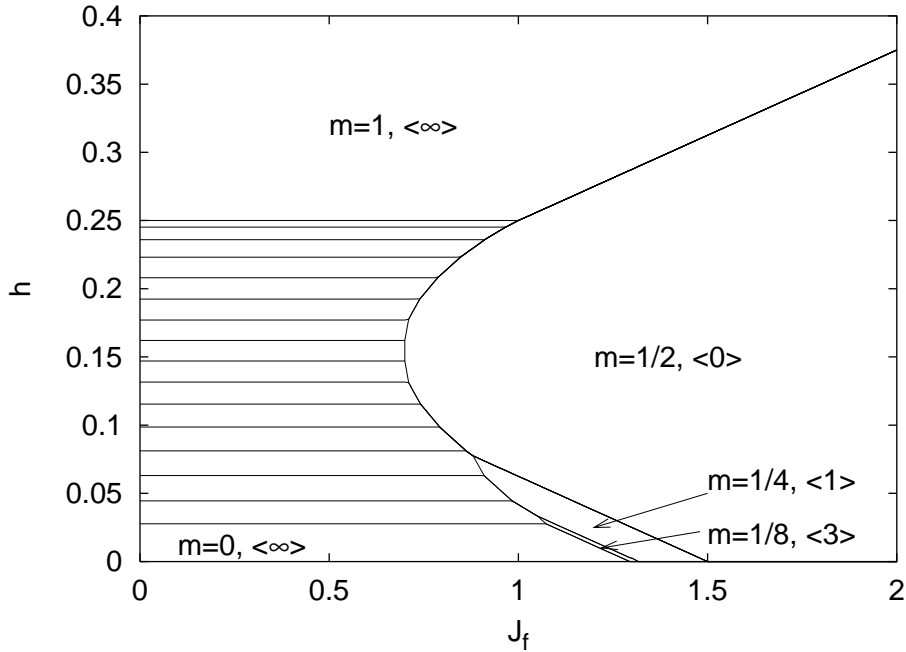,scale=1.0}
 \end{center}
\caption[Grundzustands-Phasendiagramm $J_d<0$, $h>0$]{ \label{phase_h}
 Ground-state phase diagram for the FDPC of $N=32$ spins in a
finite magnetic field $h$ with $J_p=1$ and 
large ferromagnetic $J_d \to -\infty$.
}
\end{figure}
One important difference to the antiferromagnetic case $J_d>0$ consists in the
circumstance, that for ferromagnetic $J_d$    
any  fragment of length $k$ carries a finite 
total spin  $S_{frag}=1$ in its lowest state for $h =0$, whereas the 
total spin  $S_{frag}$ for $J_d>0$ is zero for even $k$.
Hence the fragmentation is favoured  by the magnetic field, whereby short
fragments are more favourable than long fragments.      
As discussed already above, the extrema of the
energy belong to states with fragments of identical length $k$.
For $J_d <0$ in zero field the fragmented state $<\!\!k\!\!>$ ($k$ finite) 
can carry total spin  $S^z_{tot}=N/(4k+4)$.   
From the numerical inspection of the excitation spectrum of 
finite fragments we argue that their excited states with $S_{frag}>1$
are well separated from the lowest state with $S_{frag}=1$.
Hence we can expect for parameter points in the zero-field phase 
diagram (Fig. \ref{dpfmphase}) not too far from a  transition line 
that the first-order transitions can also be driven 
by an external magnetic field.

To be more precise we have studied the ground-state phases of a FDPC in a
magnetic field with
$N=32$ sites in the limit $J_d \to -\infty$.
The corresponding phase
diagram is shown in Fig. \ref{phase_h}.  
We mention, that for this chain
length the fragmented state $<\!\!2\!\!>$ does not fit to the periodic
boundary conditions and is therefore missing. (A finite-size calculation
including all zero-field ground-state phases would require $N=48$ sites
which is currently beyond the available computer facilities.) 
We argue that except for missing the $<\!\!2\!\!>$ phase 
the other ground-state phases should be well described  by the $N=32$
system, since the correlation length in all  phases  is rather small.
In particular, the transition lines between 
$m=1/4$ and $m=1/2$ and between 
$m=1/2$ and $m=1$ presented in Fig. \ref{phase_h} are exact, since these
ground-state phases belong to product states correctly described for $N=32$. 

Similar to the zero-field phase 
diagram there is no 
fragmentation for small frustration. However, supported by the field 
the fragmention sets in already 
at $J_f \approx 0.71J_p$ for $h\approx 0.13$ instead of $J_f \approx 1.30J_p$
for $h=0$. In general, the phase transitions shown in Fig.
\ref{dpfmphase} are shifted to lower values of frustration $J_f$.

\begin{figure}
 \begin{center}
  \epsfig{file=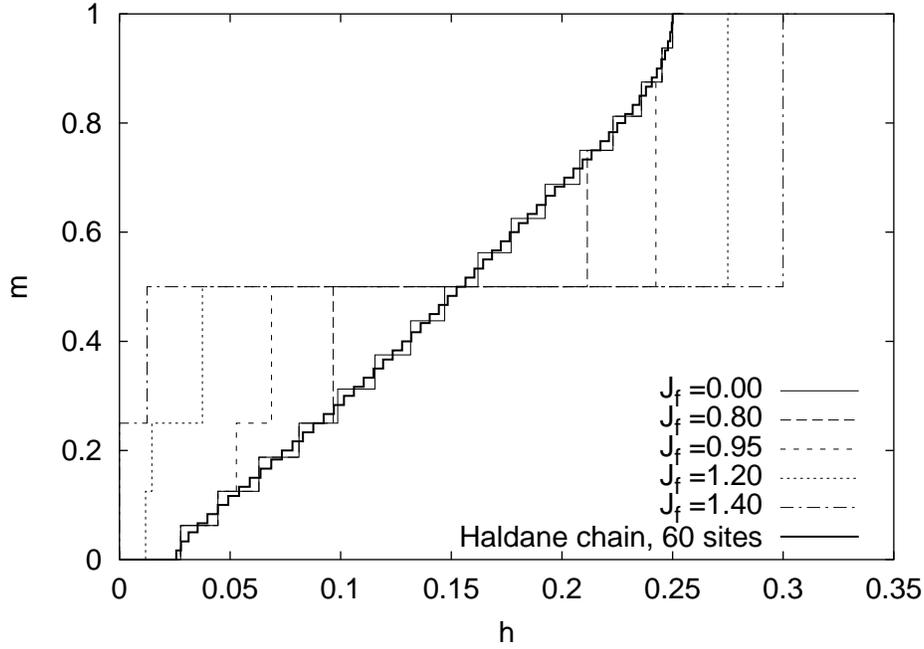,scale=1.0}
 \end{center}
\caption[magnetization curves]{ \label{m_h}
Magnetization curves $m(h)$  for
the FDPC of $N=32$ spins 
with $J_p=1$,
large ferromagnetic $J_d \to -\infty$
and various strengths of frustration $J_f$.
}
\end{figure}

The ground-state phase diagram in finite magnetic field leads to
interesting magnetization curves (Fig. \ref{m_h}).  We define the  
magnetization
as $m=2S^z_{tot}/N$, $m$ is zero in a singlet state but unity in
the fully polarized ferromagnetic state. Then the  
fragmented ground-state phases
$<\!\!\infty\!\!>$, $<\!\!3\!\!>$, $<\!\!2\!\!>$, $<\!\!1\!\!>$, 
$<\!\!0\!\!>$ in  Fig. \ref{dpfmphase} correspond 
to $m=0$, $m=1/8$, $m=1/6$, $m=1/4$ and  $m=1/2$,
respectively.

For small $J_f \lesssim 0.70 J_p$ we have a magnetization curve $m(h)$ 
typical for
unfrustrated gapped spin systems, like the Haldane chain
\cite{sakai,zhito,honecker}.
Due to the gap the curve starts with a 
plateau at zero field but then it goes continuously to saturation.
For the
finite system considered the continuous part is staircase like with small
steps (see thin solid line in Fig. \ref{m_h}) which is related to the
corresponding line of a $N=16$ Haldane chain. For comparison we added the
DMRG data for the Haldane chain of 60 sites (see thick solid line in Fig. \ref{m_h})
taken from \cite{honecker}.
For intermediate frustration  $0.71J_p \lesssim J_f \lesssim 1.08J_p$ 
we have jumps, plateaus and
continuous parts in the $m(h)$ curve.       
However, for large enough $J_f \gtrsim 1.08J_p$ we have no continuous parts in the
curve, but only plateaus connected by jumps. For instance, for $J_f=1.2J_p$
(dotted line in Fig. \ref{m_h}) we have the same sequence of 
phase transitions as in Fig. \ref{dpfmphase}, but now driven by increasing 
the magnetic field from $h=0$ to $h \approx 0.05$. 
As a result the magnetization $m$ jumps from $m=0$ to $m=1/8$ to
$m=1/6$ to $m=1/4$ to $m=1/2$ and further increasing $h$ to saturation $m=1$ 
(notice that $m=1/6$ is missing for
$N=32$ in Fig. \ref{m_h}).   
For large frustration $J_f \le 1.5J_p$  we have an extreme $m(h)$ curve 
consisting only of 
one plateau at $m=1/2$ followed by a jump to saturation $m=1$.

We remark, that the jump to the saturation $m=1$ present for 
$J_f \gtrsim 0.97J_p$
is related to independent magnon excitations versus the fully polarized
ferromagnetic state  \cite{prl02}.

We notice, that our result is in accordance with the general   
rule of Oshikawa {\it
et al.} \cite{oshikawa97} that $n(s-{\bar m})$ has to be an integer.
In our case the period $n$ of a fragmented ground state 
$<\!\!k\!\!>$ is $4k+4$, the spin
$s$ is one half and the magnetization per site ${\bar m}$ is 
$1/(4k+4)$
(the magnetization per site $\bar m$ corresponds to one half of the
magnetization $m$ used so far).
\section{Finite temperatures}
In this section we discuss some consequences of the ground-state phase
diagram on the low-temperature thermodynamics. 
The different nature of the nonfragmented phase $<\!\!\infty\!\!>$ 
and the fragmented phases $<\!\!k\!\!>$ ($k$ finite)
has an important impact on thermodynamics. While the 
the ground state $<\!\! \infty \!\!>$  is a singlet with gapped triplet
excitations, the fragmented ground states 
$<\!\! k \!\!>$ consist of independent paramagnetic entities.
As a consequence, in the former case both the susceptibility $\chi$ and the
specific heat $c$ 
are thermally activated and decay exponentially to zero if
temperature $T$ goes to zero. On the other hand, for a fragmented ground
state the susceptibility shows a paramagnetic Curie like divergency for $T
\to 0$. Hence, varying $J_f$ the low-temperature behaviour of $\chi$ changes
basically crossing the transition line between 
$<\!\! \infty \!\!>$ and $<\!\! 3 \!\!>$.
The behaviour for the specific heat near the phase boundaries 
can be more complex, since close to
these boundaries we have quasi-degenerated low-lying levels which may
lead to additional low-temperature peaks in $c$.

\begin{figure}
 \begin{center}
  \epsfig{file=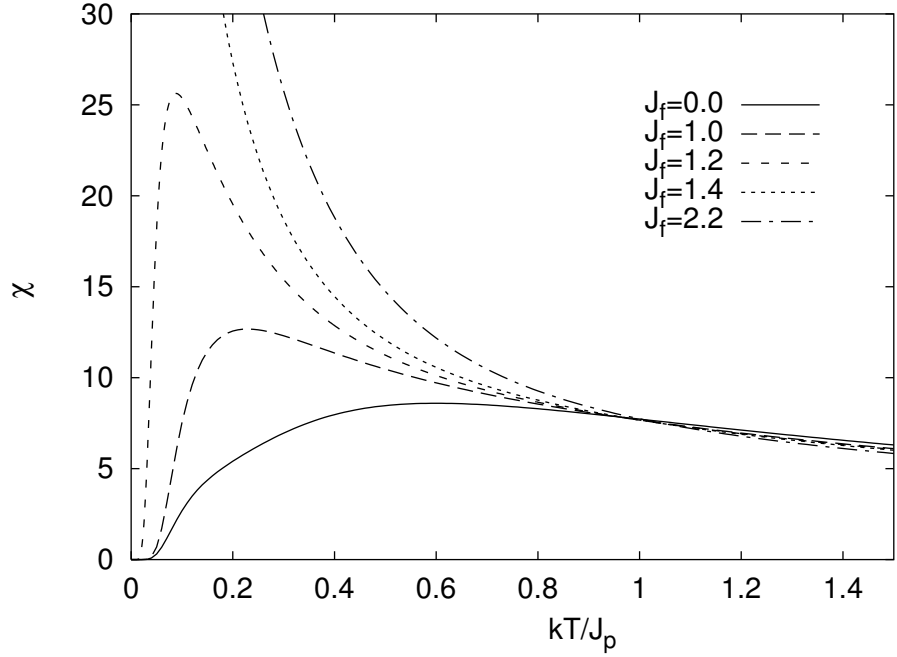,scale=0.5,angle=270}
 \end{center}
\caption[magnetization curves]{ \label{chi_T}
Magnetic susceptibility $\chi$
of the FDPC of $N=16$ spins 
with $J_p=1$,  
ferromagnetic $J_d =-400 $
and various strengths of frustration $J_f$.
}
\end{figure}

\begin{figure}
 \begin{center}
  \epsfig{file=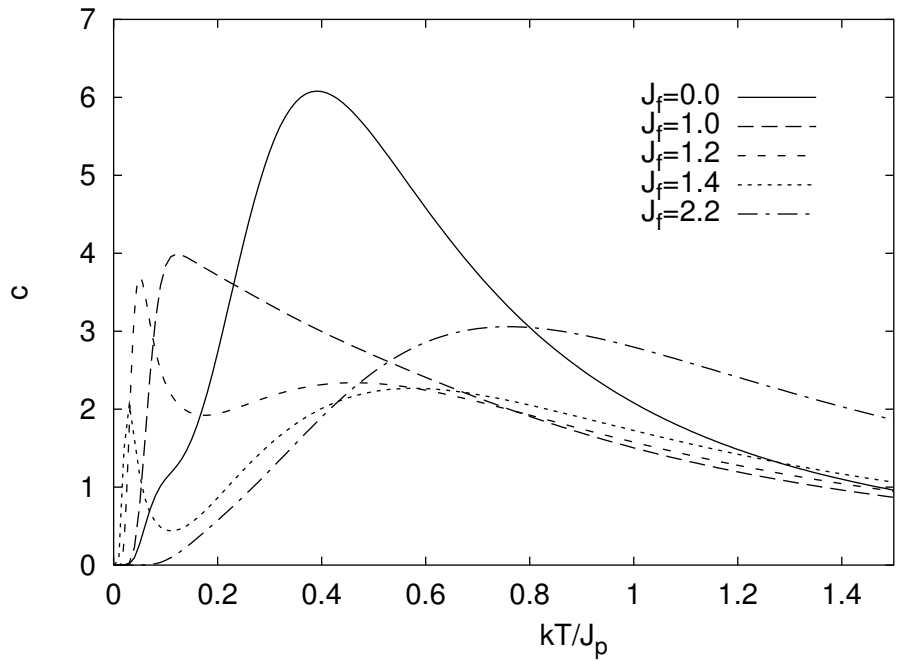,scale=0.5,angle=270}
 \end{center}
\caption[magnetization curves]{ \label{c_T}
Specific heat $c$
of the FDPC of $N=16$ spins 
with $J_p=1$,
ferromagnetic $J_d =-400 $
and various strengths of frustration $J_f$.
}
\end{figure}

To illustrate this behaviour we present in Fig. \ref{chi_T} and Fig.
\ref{c_T} the susceptibility and the specific heat for a FDPC of $N=16$
sites obtained by full diagonalization of the Hamiltonian matrix.    
The transition from spin-gap to paramagnetic behaviour is clearly seen  
in Fig. \ref{chi_T} between $J_f=1.2$ and $J_f=1.4$. In the same parameter
region the specific heat shows a characteristic double-peak structure due
to the quasi-degeneracy of states near the transition line. This effect
should survive in the thermodynamic limit.

\section{Summary}

In conclusion, we have found  a series of first-order quantum phase
transitions
in a frustrated one-dimensional quantum spin system
which is closely related to the possibility of fragmentation of the
considered chain. This series of phase transitions can be driven either by
frustration or by a magnetic field.
We emphasize that this fragmentation and the related
consequences for physical properties are pure quantum effects 
not present in classical spin systems.
The existence of different ground states has a strong impact on the
magnetization curve and the low-temperature thermodynamics of the spin
system. In dependence on the frustration the magnetization curve shows 
plateaus, jumps as well as continuous parts.

Though the possibility of fragmentation has been observed
also for frustrated two-leg spin ladder \cite{mila,honecker}, in the
spin ladder system
only very
simple ground states seem to be relevant for its magnetization curve.

The low-temperature thermodynamics can be either spin-gap like for small
frustration or paramagnetic for large frustration. Furthermore an additional
low-temperature peak in the specific heat can appear.

{\bf Acknowledgments}

This work was supported by the DFG (Ri 615/6-1). The authors are
indebted to N.B.~Ivanov for fruitful discussions and to
A.~Honecker for critical reading the manuscript and sending the DMRG data
used in Fig. \ref{m_h}.

\vspace{1cm}
%***************************************************************************
%***************************************************************************

\end{document}